# Contextual Focus: A Cognitive Explanation for the Cultural Revolution of the Middle/Upper Paleolithic

**Liane Gabora (liane@uclink.berkeley.edu)**
Department of Psychology, University of California, Berkeley
3210 Tolman Hall, Berkeley CA, 94720-1650 USA
and
Center Leo Apostel for Interdisciplinary Studies (CLEA), Free University of Brussels
Krijgskundestraat 33, B1160 Brussels, Belgium, EUROPE

**Abstract**

Many elements of culture made their first appearance in the Upper Paleolithic. Previous hypotheses put forth to explain this unprecedented burst of creativity are found wanting. Examination of the psychological basis of creativity leads to the suggestion that it resulted from the onset of *contextual focus:* the capacity to focus or defocus attention in response to the situation, thereby shifting between analytic and associative modes of thought. New ideas germinate in a defocused state in which one is receptive to the possible relevance of many dimensions of a situation. They are refined in a focused state, conducive to filtering out irrelevant dimensions and condensing relevant ones.

## Introduction: A Cultural Revolution

Human culture is widely believed to have begun between 2 and 1.5 mya, at which time a rapid increase in brain size coincides with onset of the use of fire and sophisticated stone tools. The archaeological record suggests that a perhaps even more profound cultural transition occurred between 60,000 and 30,000 ka during the Middle / Upper Paleolithic (Bar-Yosef, 1994; Klein, 1989; Mellars, 1973, 1989a, b; Mithen, 1996, 1998; Soffer, 1994; Stringer & Gamble, 1993; White, 1993). Leakey (1984) writes:

> Unlike previous eras, when stasis dominated, innovation is now the essence of culture, with change being measured in millennia rather than hundreds of millennia. Known as the Upper Paleolithic Revolution, this collective archaeological signal is unmistakable evidence of the modern human mind at work (p. 93-94).

Mithen (1996) refers to this period as the 'big bang' of human culture, claiming that it shows more innovation than the previous six million years of human evolution. It marks the beginning of a more strategic style of hunting involving specific animals at specific sites. We also see the colonization of Australia, the replacement of Levallois tool technology by blade cores in the Near East, and the first appearance of many forms of art in Europe, including naturalistic cave paintings of animals, bone and antler tools with engraved designs, ivory statues of animals and sea shells, personal decoration such as beads, pendants, and perforated animal teeth, and elaborate burial sites. Some of these items are associated with social change and the beginnings of ritualized religion; White (1982) writes of a "total restructuring of social relations" (p. 176). Moreover, we see the kind of cumulative change that Tomasello (1999) refers to as a Ratchet Effect.

What could have caused this unprecedented explosion of creativity? Some have noted that it would make things easier if this second cultural transition also coincided with an increase in brain size (Mithen, 1998; Richerson & Boyd, 2000). And in fact, human brain enlargement does seem to have occurred in two spurts. However, the second takes place between 500,000 and 200,000 (Aiello, 1996) or 600,000 and 150,000 ka (Ruff, Trinkaus, & Holliday, 1997); at any rate, well before the Upper Paleolithic. Thus the cultural revolution cannot be directly attributed to a change in the size or shape of the cranium. Leakey (1984) writes of anatomically modern human populations in the Middle East with little in the way of culture, and concludes "The link between anatomy and behavior therefore seems to break" (p. 95).

## Existing Hypotheses

Let us review some of the explanations for the Upper Paleolithic Revolution that have been put forth.

### Advent of Syntactic Language

It has been argued that while a primitive form of language, or proto-language may have existed earlier, the symbolic and syntactic aspects emerged at this time (Aiello & Dunbar, 1993; Bickerton, 1990, 1996; Dunbar, 1993, 1996). Another argument put forth is that prior to the Upper Paleolithic, language was used merely in social situations, and thereafter it became general-purpose, put to use in all kinds of situations (Aiello & Dunbar, 1993; Dunbar, 1993, 1996).

These arguments lead to the suggestion that the Upper Paleolithic Revolution was due to onset of more complex language. As noted by Tomasello (1999), language has a transformative effect on cognition. However, to posit that the cultural revolution is *due* to the attainment of sophisticated language begs the question: what cognitive change made *possible* the kind of thought that sophisticated language requires? And unless there was a underlying cognitive change involved, why did the cultural situation change so rapidly? Thus the riddle of the Upper Paleolithic is not resolved through the attainment of complex language.

### Material Culture as Externalized Memory

Donald's (1991, 2001) explanation is that material culture began to function as an externalized symbol storage. Certainly artifacts serve the purpose of anchoring knowledge or desirable states of mind so that they can be referred back to later. One sees a depiction of bison on a cave, and there is a re-living of the experience of watching a stampede of bison. One looks at notches in a log and knows how many days have gone by. The result is not so different from retrieving knowledge from memory, though the source of the knowledge is external rather than internal.

However, the external world functioned as a form of memory long before there were symbolic artifacts. A look of disapproval on a mother's face could remind a child not to eat a poisonous mushroom as readily as retrieval of a memory of doing this and getting sick. The look on the mother's face is not a material artifact, yet it functions for the child as an external memory source, in much the same way as a bison painting or notches in a log. Moreover, since material objects are manifestations of ideas, which begin in minds, it seems reasonable that we look to the mind, not the outside world, for the root cause of the cultural revolution.

### Exploration of Conceptual Spaces

Another possibility is that it reflected an enhanced ability to blend concepts (Fauconnier & Turner, 2002) or to map, explore, and transform conceptual spaces (Mithen, 1998). Mithen refers to Boden's (1990) definition of a conceptual space as a 'style of thinking—in music, sculpture, choreography, chemistry, *etc.*' As for why hominids suddenly became good at exploring and transforming conceptual spaces, he is somewhat vague:

> There is unlikely to be one single change in the human mind that enabled conceptual spaces to become explored and transformed. Although creative thinking seems to appear suddenly in human evolution, its cognitive basis had a long evolutionary history during which the three foundations evolved on largely an independent basis: a theory of mind, a capacity for language, and a complex material culture. After 50,000 years ago, these came to form the potent ingredients of a cognitive/social/material mix that did indeed lead to a creative explosion (p. 186).

Mithen may be on to something with the notion that the cultural revolution is related to the capacity to explore and transform conceptual spaces. However, although the capacity for a theory of mind, language, and complex artifacts may have, in their most primitive forms, arisen at different times, it is hard to imagine how they could have evolved independent of one another. Furthermore, if there is anything that science has established in the last decade or two it is that a single, small change in initial conditions can have enormous consequences (*e.g.* Bak, Tang, & Weisenfeld, 1988; Kauffman, 1993). Thus the possibility that the Upper Paleolithic revolution can be explained by a single change in cognitive functioning is not only the simplest explanation, it is also consistent with the sudden transition in the archeological record, and with our understanding of phase transitions across the scientific disciplines.

### Connection of Domain-Specific Modules

Many suggest that modern cognition arose through the connecting of domain-specific brain modules, stressing in particular that this would allow for the production of analogies and metaphors (Fodor, 1983; Gardner, 1983, 1993; Karmiloff-Smith, 1992; Rozin, 1976; Sperber, 1994). Mithen (1996) suggests that in the Upper Paleolithic, modules specialized to cope with domains such as natural history, technology, and social processes became connected. However, this would require that there be space enough for not just the modules but the new connections amongst them, and as we have seen, this cultural transition does not coincide with an increase in size or change of shape of the cranium. Moreover, the logistics of physically connecting these modules (whose positions in the brain evolved without foreknowledge that they should one day become connected) would be formidable. Sperber's (1994) solution is that the modules got connected not directly, but indirectly, by way of a special module, the 'module of metarepresentation' or MMR, which contains 'concepts of concepts'. However, to invent an artifact that combines information from different domains, such as an axe with a bison engraved on it, it is not necessary that the module that deals with tools be connected to the one that deals with animals, nor even that the concepts 'axe' and 'buffalo' spend time together in a meta-module. All that is necessary, as explained shortly, is that these concepts be *simultaneously accessible*.

## Psychological Basis of Creativity

We have looked at several hypotheses to account for the cultural revolution of the Upper Paleolithic. Each has merit, but none provides a satisfying explanation for this burst of creativity. To determine more precisely what gave rise to the modern human mind, it is useful to examine the psychological basis of creativity.

### Attributes of Creative Individuals

Martindale (1999) identified a cluster of attributes associated with high creativity. A first one is *defocused attention*: the tendency not to focus exclusively on the relevant aspects of a situation, but notice also seemingly irrelevant aspects (Dewing & Battye, 1971; Dykes & McGhie, 1976; Mendelsohn, 1976). A related attribute is high sensitivity (Martindale, 1977, 1999; Martindale & Armstrong, 1974), including sensitivity to subliminal impressions; stimuli that are perceived but of which one is not *conscious* of having perceived (Smith & Van de Meer, 1994).

Creative individuals also tend to have *flat associative hierarchies* (Mednick, 1962). The steepness of one's associative hierarchy is measured by comparing the number of words generated in response to stimulus words on a word association test. Those who generate few words for each stimulus have a *steep* associative hierarchy, whereas those who generate many have a *flat* associative hierarchy. Thus, once such an individual has run out of the more usual associations (*e.g.* 'chair' in response to 'table'), unusual ones (*e.g.* 'elbow' in response to 'table') come to mind. The evidence that creativity is associated with both defocused attention and flat associative hierarchies suggests that creative individuals not only notice details others miss, but these details get stored in memory and are available later on.

However, a considerable body of research suggests that creativity involves not just the ability to defocus and free-associate, but also the ability to focus and concentrate (Barron, 1963; Eysenck, 1995; Feist, 1999; Fodor, 1995; Richards et al. 1988; Russ, 1993)[1]. As Feist (1999) puts it: "It is not unbridled psychoticism that is most strongly associated with creativity, but psychoticism tempered by high ego strength or ego control. Paradoxically, creative people appear to be simultaneously very labile and mutable and yet can be rather controlled and stable" (p. 288). He notes that, as Barron (1963) put it: "The creative genius may be at once naïve and knowledgeable, being at home equally to primitive symbolism and rigorous logic. He is both more primitive and more cultured, more destructive and more constructive, occasionally crazier yet adamantly saner than the average person" (p. 224).

### Phases of the Creative Process

How do we make sense of this seemingly paradoxical description of the creative individual? The evidence that creativity is associated with both defocused free-association and focused concentration is in fact consistent with the idea that the creative process consists of a *generative* phase followed by an *evaluative phas*e (Boden, 1991; Dennett, 1978). Indeed there is an enduring notion that there are two kinds of thought, or that thought varies along a continuum between two extremes (Ashby & Ell, 2002; James, 1890/1950, Johnson-Laird, 1983; Neisser, 1963; Piaget, 1926; Rips, 2001; Sloman, 1996). Although the issue is still a subject of hot debate, the general picture emerging is as follows. At one end of the continuum is an intuitive, *associative mode* conducive to finding remote or subtle connections between items that are *correlated* but not necessarily *causally* related. This mode may yield an idea or problem solution, though perhaps in a vague, unpolished form. At the other end of the continuum is a rule-based, *analytic mode* of thought, conducive to analyzing relationships of *cause and effect*. This mode facilitates fine-tuning and manifestation of the creative work.

## Contextual Focus Hypothesis

Let us now now look at a tentative explanation of the cognitive mechanisms underlying the creative process (Gabora, 2000, 2002a, 2002b; Gabora & Aerts, 2002). We take as a starting point some fairly well-established features of memory. According to the doctrine of *neural re-entrance*, the same memory locations get used again and again (Edelman, 1987). Each memory location is sensitive to a range of *subsymbolic microfeatures* (Smolensky, 1988), or values of them (Churchland & Sejnowski, 1992). Location *A* may respond preferentially to lines of a certain angle (say 90 degrees), neighboring location *B* respond preferentially to lines of a slightly different angle (say 91 degrees), and so forth. However, although location *A* responds *maximally* to lines of 90 degrees, it responds to a lesser degree to lines of 91 degrees. This kind of organization is referred to as *coarse coding*. The upshot is that storage of an item is *distributed* across a cell assembly that contains many locations, and likewise, each location participates in the storage of many items (Hinton, McClelland, & Rummelhart, 1986). Items stored in overlapping regions are correlated, or share features. Therefore memory is *content addressable;* there is a systematic relationship between the state of an input and the place it gets stored. Thus episodes stored in memory can thereafter be evoked by stimuli that are similar or 'resonant' (Hebb, 1949; Marr, 1969).

Let us consider the significance of this memory architecture for creativity. To be constantly in a state of defocused attention, in which relevant dimensions of a situation do not stand out strongly from irrelevant ones, would be clearly impractical. It is only when one does not yet know what *are* the relevant dimensions—or when those assumed to be relevant turn out not to be—that defocused attention is of use. After the relevant dimensions have been found it is most efficient to focus on them exclusively. Indeed it has been shown that in stimulus classification tasks, psychological space is stretched along dimensions that are useful for distinguishing members of different categories, and shrunk along nonpredictive dimensions (Nosofsky, 1987; Kruschke, 1993). In ALCOVE, a computer model of category learning, only when activation of each input unit was multiplied by an attentional gain factor did the output match the behavior of human subjects (Kruschke, 1992; Nosofsky & Kruschke, 1992). Thus learning and problem solving involve both (1) associating stimuli with outcomes, and (2) shifts in attention that determine how one 'parses' the space. Let us refer to the situation in which many stimulus dimensions or aspects of a situation activate memory to an almost equal degree as a flat activation function, and the situation in one focuses exclusively on one stimulus dimension, or aspect of a situation, as a spiky activation function. We can refer to the ability to spontaneously adjust the shape of the activation function in response to the situation at hand as the capacity for *contextual focus*.

[1] There is also evidence of an association between creativity and high variability in physiological measures of arousal such as heart rate (Bowers & Keeling, 1971), spontaneous galvanic skin response (Martindale, 1977), and EEG alpha amplitude (Martindale & Hasenfus, 1978; Martindale, 1999).

Let us now explore the possibility that creative individuals are not always in a state of defocused attention, but that they can enter this state when useful (such as in a word association test). Thus when one encounters a problem or inconsistency, or seeks self-expression, one enters a state of defocused attention conducive to associative thought through a flattening of the activation function. More diverse memory locations get activated and provide ingredients for the *next* thought. Because of the distributed, content-addressable structure of memory, a seemingly irrelevant element of the situation may evoke an episode from memory that shares this element. The connection between them may inspire a new idea.

The vague idea generated in a defocused state is clarified by focusing attention on the new connection and shifting to a more analytic mode. The activation function becomes spikier, and the region searched and retrieved from narrower. This continues until, to use Posner's (1964) terms, one has filtered out the irrelevant dimensions and condensed the relevant ones.

In sum, it is proposed that the capacity for contextual focus is the distinguishing feature of the modern human mind, and the reason for the cultural revolution of the Upper Paleolithic.

### Why Connected Modules is Not Necessary

Let us see why to blend items from different modules together it is not necessary that they be connected. Once modules start encoding more of the richness of their respective domains, to the extent that these domains have elements in common, an individual will start to have experiences that activate multiple modules simultaneously. Let us say, for instance, that Mithen is right about natural history and social situations being the domain of different modules. Both these modules will come to contain memory locations that respond to the chaotic, volatile elements of a situation. So, for example, an individual in a volatile mood might remind one of stormy weather. Thus the modules themselves need not be connected for a blend to occur so long as a situation can simultaneously activate them.

### Possible Explanation for Lag Between Brain Expansion and Cultural Revolution

An increase in brain size provides more storage space, thus memories can be laid down in richer detail. But it doesn't follow that this increased space could immediately be navigated in the most efficient way. This is particularly the case for situations involving simultaneous activation of multiple modules; there is no reason to think their contents would be compatible enough to coexist in a stream of thought. The stormy individual might prompt one to take shelter without leading to the without realization that ones' own mood could take on this stormy quality. A situation where three apples reminds one of ones' three children could prompt one to bring home the right number of apples, without leading to the realization that 'three-ness' can be independent of apples or children. It seems reasonable that it took time to fine-tune the cognitive system such that items from different domains could be blended together and recursively redescribed in a coordinated manner. Only then could the full potential of this large brain be realized.

## Summary and Discussion

The period of history that exhibits the most impressive cultural transition is the Upper Paleolithic. To gain insight into what caused this unprecedented explosion of creativity, it is useful to examine the creative process. Creative individuals are prone to states of defocused attention, and tend toward flat associative hierarchies, suggesting a proclivity for associative thought. However, creativity involves not just an intuitive, associative mode of thought, but also an analytic, evaluative mode. This suggests that creativity requires the ability to shift between these modes.

Thus it is tentatively proposed that the arrival of art, science, religion, and likely also complex language and a restructuring of social relationships in the Upper Paleolithic was due to the onset of contextual focus: the capacity to focus or defocus attention in response to the situation, thereby shifting between analytic and associative modes of thought. New ideas germinate in a defocused state in which one is receptive to the possible relevance of many dimensions of a situation. They are refined in a focused state, conducive to filtering out irrelevant dimensions and condensing relevant ones.

Contextual focus is not a matter of more memory, but of a more sophisticated way of *using* memory; thus the proposal is consistent with there being no increase in brain size at this time. It is consistent with a point made by Bickerton (1990) and Leakey (1984) that brain size cannot be equated with intelligence. However it deviates from Bickerton's perspective in that it is not language *per se* that made the difference, but rather a kind of cognitive functioning that made not only language possible, but all aspects of group survival that can benefit from being considered from different perspectives and at different degrees of abstraction. It does not require that modules be connected, but merely simultaneously accessible.

## Acknowledgments

I would like to thank Ellen Van Keer for comments. This research was supported by Grant G.0339.02 of the Flemish Fund for Scientific Research.